\documentstyle[12pt]{article}
\title{On Hyperbolic ``Theories" of Relativistic
   Dissipative Fluids}
\author{Robert Geroch\thanks{E--mail: geroch@midway.uchicago.edu}
\\Enrico Fermi Institute\\5640 Ellis Ave, Chicago, IL - 60637}
\begin{document}
\maketitle

\noindent{\bf Abstract.} It is argued that, at least for the case of
Navier-Stokes fluids, the so-called hyperbolic theories of dissipation
are not viable.
\\

Consider a Navier-Stokes fluid.  By this, we mean a substance
that, within a certain regime\footnote{That is, in the limit as one
moves further and further into that regime.  More on this later.},
may be described as follows.  First, the local state of the substance 
must be completely characterized, within this regime, by
the values of precisely five fields on
space-time: two scalar fields, $\rho$ and $n$;
two vector fields, $u^a$ and $q^a$, the former unit timelike
and the latter orthogonal to $u^a$; and one tensor field,
$\tau^{ab}$, symmetric and also orthogonal to $u^a$.  These fields
are interpreted, respectively, as the fluid mass density,
particle-number density, velocity, heat flow and stress.   Second, the 
substance must be found to behave, within this regime, such that the 
following system of four equations \cite{E71} holds:
\begin{equation}
\nabla_b((\rho + p)u^au^b + pg^{ab} + 2 q^{(a}u^{b)} + \tau^{ab}) = 0,
\label{NS1}
\end{equation}\begin{equation}
\nabla_m(nu^m) = 0,
\label{NS2}
\end{equation}\begin{equation}
q_a + \sigma \rho (\delta_a{}^b + u_au^b)
[(\nabla_b T)/T + u^m\nabla_m u_b] = 0,
\label{NS3}
\end{equation}\begin{equation}
\tau_{ab} + \eta \rho (\delta_{(a}{}^m + u_{(a}u^m)
\nabla_{|m|}u_{b)} + (\xi - \eta/3) \rho (g_{ab}+u_au_b)\nabla_mu^m = 0.
\label{NS4}
\end{equation}
The five new variables, $p, T, \sigma, \eta$, and $\xi$, appearing
in these equations are fixed state functions, i.e., they 
are functions of $\rho$ and $n$, to be specified, once and for all, 
at the beginning.  These five variables are interpreted, respectively, as the
fluid pressure, temperature, thermal conductivity, viscosity and
bulk viscosity.  Eqn (\ref{NS1}) is interpreted as conservation
of stress-energy, and Eqn. (\ref{NS2}) as conservation of particle number.
Eqn. (\ref{NS3}) is interpreted as describing heat-flow,
and (\ref{NS4}) as describing viscosity.  

Examples of Navier-Stokes fluids include ordinary liquids, such as water, 
and ordinary gases, such as nitrogen at standard temperature and pressure.  

We shall here be concerned with the issue of what happens outside the
regime in which the Navier-Stokes description of such a
fluid is applicable.  In particular, we shall be concerned with whether 
or not there are available other physical theories that apply
when the Navier-Stokes description breaks down.  

Note that we are concerned {\em solely} with this single 
class of substances --- Navier-Stokes fluids.  There 
are, certainly, many other types of substances in the world ---  some
characterized by more fields, and some simply by different 
fields\footnote{An example of a substance requiring more fields is Helium 2,
the fluid that manifests ``second sound".  See, e.g., \cite{ZG71}.
For other examples of non-Navier-Stokes substances, see \cite{APR98}
and references therein.}.   
Undoubtedly, many of these substances manifest dissipative effects.  It is 
certainly conceivable that some of these substances, in their dissipative 
regimes, may be described by viable hyperbolic theories.   But the variety 
of possible ``substances" in this world is enormous.
We have to start somewhere, and the class of Navier-Stokes fluids seems as 
good a starting point as any:  These fluids are rather common,
are relatively well-understood, and are simple to describe 
mathematically.  Furthermore, these fluids --- or, rather, the
breakdown of the Navier-Stokes descriptions of their dissipation ---
represent relatively pure examples of the descent
into molecular effects.  Perhaps, ultimately, it will  be possible to extend
some of these considerations to other types of substances.
 
Fix a Navier-Stokes fluid.  The breakdown of the Navier-Stokes
description of this fluid typically takes place at small distances,
say, at all distances less than the order of some distance scale $L$. In
general, this ``breakdown" takes two distinct forms.  One the one hand,
the Navier-Stokes equations, (especially (\ref{NS3})-(\ref{NS4})), 
may fail.  That is, the expressions on the left in these equations
may be found to be nonzero, e.g., by an amount the order of the 
sizes of the terms in these equations.  Indeed, relativity requires 
failure of the system (\ref{NS1})-(\ref{NS4}) at least at distances 
of the order of $d/c$, where $d$ denotes the value of
a typical Navier-Stokes dissipation coefficient ($\sigma, \eta$, or
$\xi$), and $c$ the speed of light. 
The second --- and potentially more serious --- type of breakdown
involves ambiguity as to the physical meanings of the
variables that appear in this description --- i.e., of the quantities
$\rho, n, u^a, q^a, \tau^{ab}, p, T$, etc.  
To illustrate this, let the Navier-Stokes fluid be nitrogen gas, and
consider a sample of this fluid, at approximately standard temperature 
and pressure, for which changes in fluid state take place only over 
distances of the order of centimeters.  For this sample, we know, more 
or less, what fluid velocity and temperature mean, i.e., we could
construct centimeter-sized probes to measure them.  But now suppose that,
instead, our sample manifested state-changes taking place
on distance scales of the order of $10^{-6}$ cm, the
mean free path of the nitrogen molecules. 
In this case, things are not so simple:  We must construct probes
that work on these smaller scales.  What typically happens is
that different types of probes --- although they all give the same
result on centimeter-scales --- will begin to give different results
on the scale of $10^{-6}$ cm.  ``Velocity" and ``temperature", on
scales comparable with $L$, become ambiguous.  

This $L$ is a state function, i.e., is a function of $\rho$ and $n$.  
We remark that inherent in this basic notion of a breakdown scale
are at least two potential complexities.  First, it is possible, at
least in principle, that the two Navier-Stokes equations, 
(\ref{NS3}) and (\ref{NS4}), could fail on quite different scales;
or that the various Navier-Stokes variables could lose their meanings
on quite different scales.  These circumstances would
necessitate replacing the single scale $L$ by a variety of different
scales.  We shall return to this possibility later.  A second type
of complexity involves the possibility that the breakdown of an
equation or a variable occurs in a more complicated manner --- one
that cannot be described simply by a length.
For example, the requirement imposed by relativity on the system
(\ref{NS1})-(\ref{NS4}) is, in more detail, the following:  
This system of equations must fail on every combined distance-time
scale, $L, t$, satisfying $L^2 \stackrel{<}{\sim} dt$ and $L > ct$, where 
$d$ denotes the value of a typical Navier-Stokes dissipation coefficient.  
These two inequalities describe some region of the $L,t$-plane\footnote
{The complement of this region, incidentally, includes arbitrary
small $L$-values and arbitrarily small $t$-values!}.
While these two types of complexities are certainly possible in
principle, neither seems to arise in simple examples.
For nitrogen gas at standard temperature and pressure, for instance,
the Navier-Stokes equations and variables all break
down on the same scale, $L = 10^{-6}$ cm, which is also the order of
$d/v$, where $d$ is a typical Navier-Stokes dissipation coefficient,
and $v$ is the speed of sound. 

Fix a Navier-Stokes fluid, with breakdown scale $L$.  There is a class
of systems of partial differential equations, the so-called
``hyperbolic theories", that purport to be applicable to such a fluid
on scales of the order of, and smaller than, $L$.  These systems
\cite{LMR86} \cite{GL90} \cite{MR93}  have 
the following structure.  Their fields consist of the same five fields 
that appear in the basic Navier-Stokes system, namely $(\rho, n,
u^a, q^a, \tau^{ab})$.  But their equations are different.  
The conservation equations, (\ref{NS1}) and (\ref{NS2}), 
are retained; while the dissipation equations, (\ref{NS3}) and (\ref{NS4}), 
are replaced by two new equations of the same tensor character (i.e., one 
a $u$-orthogonal vector and the other a $u$-orthogonal symmetric tensor).   
These two new equations can involve the values of all the variables,
together with, linearly, the values of all their first derivatives.
This entire quasilinear, first-order system of partial differential
equations is so structured that it has two key properties:  i) The
system is symmetric-hyperbolic, and so in particular it has (unlike the
original Navier-Stokes system (\ref{NS1})-(\ref{NS4})) an initial-value
formulation; and ii) the system reproduces in the appropriate regime,
i.e., on scales much larger than $L$, the original Navier-Stokes system. 
We remark that there is a vast number of possible candidates for this
system of partial differential equations.  For example, the general
first-order, quasilinear, symmetric-hyperbolic system of partial differential 
equations in the variables $(\rho, n, u^a, q^a, \tau^{ab})$, manifesting the 
conservation laws (\ref{NS1})-(\ref{NS2}), involves 219 free (scalar) 
functions of 8 (scalar) variables\footnote{Is it possible to reduce this 
plethora of possibilities by identifying some small, natural class of simple
hyperbolic extensions of the system (\ref{NS1})-(\ref{NS4})?  In this
connection, we note that the obvious extension \cite{HP01} --- the result
of adding multiples of $(g_{as} + u_au_s)u^m\nabla_m q^s$ and 
$(g_{as} + u_au_s)(g_{bt} + u_bu_t)u^m\nabla_m \tau^{st}$
to the left sides of Eqns. (\ref{NS3}) and (\ref{NS4}), 
respectively --- fails, for the resulting system of equations is not, 
apparently, symmetric-hyperbolic.}!  

To what extent does such a system of partial differential equations 
represent a viable ``theory" for a Navier-Stokes fluid, applicable at 
scales approaching and less than the breakdown scale $L$?
There arises an immediate difficulty:  The variables
$(\rho, n, u^a, q^a, \tau^{ab})$, in terms of which this new
system of equations is written, become physically ambiguous
on the very distance scale, $L$, at which this system 
purports to be applicable.   That is, as matters stand there are
no meaningful ``quantities" about which this new theory could
make assertions.  There is, of course, a simple cure for this malady:  
Include, as part of the statement of the new theory, a detailed
description of what experimental procedures are to be used,
on scales comparable with and smaller than $L$, to observe its
variables\footnote{In the case of a gas, for example, such a description
might entail writing down formulae, in terms of the distribution
function, for these variables, leaving to the experimentalist the
task of designing the appropriate tiny instruments to reflect these formulae.}.
But even supposing for a moment that such a description has been
provided, there arises a second difficulty.  
Inherent in the breakdown of the physical meanings of the five variables
$(\rho, n, u^a, q^a, \tau^{ab})$ is the inability of these variables
--- even as reinterpreted within the new theory --- to 
provide a complete local description of the fluid state.
That is, this variety of new potential observables arising on scales
of the order of $L$ implies that there is more happening within
the fluid at these scales than can be described by {\em any} five 
variables. As a consequence, we could hardly expect any viable system 
of fluid equations involving just the original variables 
$(\rho, n, u^a, q^a, \tau^{ab})$. 

It could be argued that this new system of equations should be regarded as 
merely the ``second-order"  description of an actual physical fluid\footnote
{It may be that not all of the possible hyperbolic systems of equations 
--- involving all 219 functions of 8 variables ---
would be admitted as viable ``second-order" systems.  
Unfortunately, it is not clear which systems would and which 
would not be admitted, because it is not clear 
exactly how this expansion in ``orders" is to be carried out.  For example,
would there be admitted in the second-order system a term quartic 
in the heat-flow $q^a$, provided it is preceded by a sufficiently large 
coefficient?}.  The ``zeroth-order" description is as a classical
perfect fluid, with equations (\ref{NS1})-(\ref{NS2}) (conservation); 
and the ``first-order" description is as a Navier-Stokes fluid, with 
equations (\ref{NS1})-(\ref{NS4}).  Presumably there will
ultimately be written down third- and then still higher-order 
descriptions --- systems involving additional equations, and, very likely, 
also additional fields.  This view is in fact supported by certain 
calculations in statistical mechanics \cite{HG49} \cite{JS71} \cite{WIJS79}.  
Consider, specifically, a gas described by a distribution function,
subject to Boltzmann's equation with some appropriate collision
function.  Introduce the various moments of the distribution function,
and interpret these moments as fields (infinite in number), which, 
taken together, describe the local properties of the fluid.  Next,
using the Boltzmann equation, obtain the system of equations (also infinite
in number) connecting these moments. 
What is found, of course, is that the resulting system never ``closes", 
i.e., higher-order moments continue to exert their effects on
the space-time behavior of lower-order moments.  But we may,
nevertheless, truncate this system:  Assign to these various moments
orders, and then write down, for each order, a system of equations involving
only the moments up to and including those at that order.  We remark that
the second-order system of equations obtained in this way may or may not
turn out to be symmetric-hyperbolic, depending on the truncation 
scheme employed.  Indeed, there is no reason either to expect or
to demand  that a resulting second-order system 
 --- or any of the systems obtained at higher order --- will 
have any solutions at all\footnote{Of course, we do expect that the entire
coupled system, including the terms of all orders, will have an appropriate 
class of solutions, since we know that this is the case for the 
Boltzmann equation itself.}.

Consider, then, such a hierarchy of systems of equations --- of successive 
orders zero, one, two, etc --- that come into play on scales comparable
with $L$.  To what extent should the second-order system in this
hierarchy be regarded as a viable physical theory?   Clearly, there
is a potential problem here, for it is difficult to isolate, and thus 
construct a theory of, the pure ``second-order effects":  Just as these 
effects  come into play, so do interfering effects from higher orders.
To illustrate this point, let us consider again the original 
Navier-Stokes system, (\ref{NS1})-(\ref{NS4}), i.e., the ``first-order"
system.  We observe --- and this is a key point --- that this system
differs in an essential way from those of all higher orders.  
Navier-Stokes effects can survive to large distance scales, in a way that
higher-order effects cannot.  Take, for example, 
Navier-Stokes Eqn. (\ref{NS3}).  Consider the following
experimental arrangement, in the indicated limit.  
Let the temperature gradient be very much smaller than $T/L$,
let this gradient be maintained over a distance very much larger
than $L$, and, finally, wait a very long time (to allow
the resulting heat-flow to produce a significant effect on the fluid 
states at the two ends of this temperature-gradient).   In this limit,
the observed effects of thermal conductivity, as described by
(\ref{NS3}), remain finite, while the effects of any ambiguities
as to the meanings of the variables, together with any higher-order 
effects, approach zero.  The Navier-Stokes system, in other words, has 
a ``regime of applicability" --- a limiting circumstance in which the 
effects included within that system remain prominent while the effects 
not included become vanishingly small.  The Navier-Stokes system, 
in short, is essentially a part of continuum mechanics\footnote
{This feature may be related to the following observation. 
What is actually observed about a fluid, on macroscopic scales,
is typically only its state functions.  Heat-flow and stress are then
inferred from these, e.g., in the case of heat-flow, by observing changes
in the state of the fluid residing at either end of a temperature gradient.  
In other words, we infer heat-flow and stress through failure of conservation 
of the naive stress-energy, $(\rho + p)u^au^b + p g^{ab}$.  
But there is only so much information that can be extracted from the failure 
of this one conservation law, i.e., there are only so many variables that 
can be inferred in this way.  This window seems to be exhausted already by 
$q^a$ and $\tau^{ab}$. Is there some precise result along these lines?}
 --- it was (or at least, could have been) discovered before we even knew 
about atoms; it is the system one uses, e.g., to decide which
coat to wear in the morning  --- while the higher-order systems
are essentially microscopic in character.   Indeed, it is a reflection
of this feature that there is any  sensible notion at all of 
a ``Navier-Stokes fluid", as described earlier. 

There is no similar regime of applicability for
the second-order systems of equations:  Just as the effects
described by a second-order system begin to become significant,
those effects are confounded
by ambiguities as to the meanings of variables and by higher-order
effects.  In connection with this point, we remark that there has not 
been given, as far as I am aware, any example of a hypothetical experimental
result such that an actual observation of that result would be interpreted 
as ruling out the second-order ``theories" for Navier-Stokes fluids.   
The second-order systems of equations, in short, cannot stand alone as  
physical theories.  The second- and higher-order systems of equations
for a Navier-Stokes fluid might thus be compared with attempts to find
equations of motion for extended bodies in general relativity
\cite{MTW73} (an area plagued by problems of the meanings of variables),
or with the ppn parameterization \cite{CW79} for gravitational theories 
(in which there is an ample supply of parameters to fit observations).
In neither of these two cases would the system of equations, truncated
at the second order, be regarded as a viable physical theory.

We remarked earlier that it is possible, at least in principle, 
that a Navier-Stokes fluid could have the feature that its
equations, (\ref{NS3})-(\ref{NS4}), break down on some distance scale much
larger than that on which the meanings of its variables,
$(\rho, n, u^a, q^a, \tau^{ab})$, break down\footnote{The reverse,
of course, is impossible:  The equations could hardly make sense
when the variables in those equations do not.}.  This is an
intriguing possibility, for it would represent a fertile ground \cite{HP01}
for a viable hyperbolic theory:  Such a fluid would manifest an
intermediate regime, in which the variables $(\rho, n, u^a, q^a, \tau^{ab})$ 
are meaningful, and are at the same time free to satisfy
some system of equations other than Navier-Stokes\footnote 
{We remark on a second possibility.  There 
might exist a substance that is characterized by precisely the Navier-Stokes
variables, $(\rho, n, u^a, q^a, \tau^{ab})$, but which fails to 
satisfy Eqns. (\ref{NS3})-(\ref{NS4}) in any regime, even on large scales.
Such a substance would not, of course, be a Navier-Stokes fluid as we have
characterized it.   Yet, if any such substance exists, it too would 
be a candidate for having a description by a viable hyperbolic theory.}.
Unfortunately, I am aware of no Navier-Stokes fluid --- either an actual 
one or a gedanken fluid --- having this character.  Here, as an example, 
is one attempt to construct such a fluid.  Consider a gas consisting
of a mixture of approximately equal particle-numbers of
helium atoms and neutrinos, at approximately standard temperature
and pressure.   [Assume, for present purposes,
that these particles interact only by elastic scattering.]
One would expect that the helium atoms (since they contribute the bulk
of the mass of the fluid) will control the basic fluid behavior, 
while the neutrinos (since they have the much larger mean free path)
will control dissipative effects.  Unfortunately, this particular 
substance is not a Navier-Stokes fluid in our sense,
for to characterize its local state requires,
in addition to the five Navier-Stokes variables, a sixth variable:
the relative particle numbers of the two species.  But it is not
difficult to extend the Navier-Stokes system so that it is
applicable to this fluid:  Include within the system this
additional variable, together with an additional (diffusion)
equation to describe its behavior\footnote
{It is apparently not known whether there exists {\em any}
symmetric-hyperbolic system of equations appropriate to this extended
Navier-Stokes system.  Probably, there does.}.  In any case, here 
is an example of a fluid having two characteristic distance scales:  
one large (the neutrino mean free path) and one small (the helium-atom 
mean free path).  One could imagine that the Navier-Stokes equations 
might fail on the larger scale while the meanings of the variables 
fail on the smaller.  But, unfortunately, things do not work out this way:
In this example, the meanings of the basic Navier-Stokes variables break
down already on the larger distance scale.  For instance, take a sample 
of this fluid whose state changes on distance scales smaller than the
neutrino mean free path.  Then the  
result of a temperature measurement on this distance scale will depend on
how opaque one's thermometer is to neutrinos; and a measurement
of the ``fluid velocity" will depend on how helium atoms and neutrinos
are counted as to their contributions to this velocity.  That is, there
is a variety of notions of ``temperature" and ``velocity" on this scale,
notions that, while they all agree with each other on scales much larger
than the neutrino mean free path, do not on these smaller scales.  This 
particular fluid, in short, does not present us with any intermediate 
regime that would be appropriate for a second-order
system of equations.  It would be interesting to find
some example of a fluid manifesting such an intermediate regime --- or 
to produce a good argument that no such fluid exists.   

As we remarked earlier, the Navier-Stokes system of
equations, (\ref{NS1})-(\ref{NS4}), must, for any given fluid, fail 
on some sufficiently small distance scale $L$; and we could, of course, 
design a series of experiments, carried out on that fluid near that scale, 
to observe such discrepancies.
We might then proceed to ``interpret" these experiments in terms of the
second-order systems of equations, in the following manner.  First,
we must set up some correspondence between the various quantities that have
been observed in these experiments and the variables appearing in the 
second-order systems, for no such correspondence is currently provided
by these systems.  Having done this, we may use these experimental
results to assign values to some of the free parameters or functions
that appear in the second-order systems.  The result of this will
be a second-order system of equations that ``fits the experiments
better than does the Navier-Stokes system" (that is, 
does so for this particular series of experiments, under this particular 
choice of a correspondence between the observed quantities and the variables
that appear in the equations).  But it is difficult to argue that such a fit 
constitutes either experimental evidence
in favor of these second-order systems or evidence that these 
systems of equations are viable physical theories.  The second-order
systems are sufficiently encompassing that one or another of these systems 
is available to accommodate virtually any ``effect" arising at scale
$L$, including, not surprisingly, whatever effects happen to arise,
in this particular fluid, in connection with the failure of 
its Navier-Stokes description. 

Then how, if at all, are we to understand Navier-Stokes fluids in relativity? 
On the face of it, we seem to have no theory at all in the usual sense\footnote
{Of course, there always remains the ``theory" that carries the full local
complexity of the fluid on small distance scales, i.e., the
theory that, for the case of a gas, carries the entire distribution function.  
But this description requires an infinite number of local fields.}:
The Navier-Stokes description must (by relativity) fail
on small distance scales; while the second-order systems of equations
arguably do not qualify as physical theories.  First note that we do
indeed have a ``theory" of a somewhat different sort.  Fix a physical 
Navier-Stokes fluid, and consider the entire panoply of second-order 
systems of partial differential equations appropriate for that
fluid, i.e., consider all those systems, in the variables
$(\rho, n, u^a, q^a, \tau^{ab})$, that are symmetric-hyperbolic
and have the correct Navier-Stokes limit.  All of these systems 
predict \cite{RG95} that, on large distance scales, the left
sides of Eqns. (\ref{NS3})-({\ref{NS4}) will remain
small throughout time, i.e., they predict ``Navier-Stokes behavior".   
Furthermore, all of these systems have initial-value formulations.   This 
guarantees that they can be used to solve physical problems, e.g., that they 
could be combined with Einstein's equation to study stability
of gravitating fluid systems.  It is futile
to debate which of these second-order systems is the ``correct" one.
In the regime in which they differ significantly from each other,
the systems themselves break down \cite{RG95}:
Their variables lose their meanings, and higher-order
effects come into play.  We may think of this situation in the
following way.  On small distance scales, we have no suitable
theory at all.  On larger distance scales, by contrast, there is one
system of equations --- the Navier-Stokes system --- that is
appropriate for the description of the physics of the fluid, and
a second family of systems --- the hyperbolic systems --- that are
appropriate for the mathematics.  This splitting of the physics and 
the mathematics is a novel situation, and it takes some getting used to.  
But, with a little care, ``theories" of this type can be applied as 
effectively as more traditional physical theories.

It might be of interest to carry out a careful analysis of various
other substances \cite{APR98}, to see whether, in any of these
cases, there can be recovered some remnant of a viable ``hyperbolic 
theory of dissipation".  It would be of particular interest to find an
example of a substance for which it is possible 
i) to spell out the (limiting) regime in which the effects of the
theory will be predominant, ii) to provide a complete list of the 
variables describing that substance, i.e., of those variables that 
characterize the local behavior of the substance in that regime, 
iii) to indicate how those variables are to be measured in that regime,
iv) to write out, in full, the system of equations imposed by the
theory on those variables, indicating any remaining free functions or
parameters, and v) to give some credible evidence, experimental or 
theoretical, that those variables, in that regime, actually satisfy 
those equations.

I wish to thank Oscar Reula, Gabriel Nagy, and Rafael Sorkin
for a number of helpful comments and suggestions.


\begin{thebibliography}{99}


\bibitem{APR98} A.M. Anile, D. Pavon, V. Romano, ``The Case for Hyperbolic
Theories of Dissipation in Relativistic Fluids" (1998), gr-qc/9810014

\bibitem{E71} C. Eckert, Phys Rev 58, 919 (1940)

\bibitem{ZG71} Z. M. Galasiewicz, ``Helium 4" (Pergamon Press, NY, 1971),
pp 19.

\bibitem{GL90} R. Geroch, L. Lindblom, Phys Rev D 41, 1855 (1990)

\bibitem{RG95} R. Geroch, J. Math. Phys. 36, 4226 (1995)

\bibitem{HG49} H. Grad, Comm Pure Appl Math 2, 331 (1949)

\bibitem{HP01} L. Herrera, D. Pavon, ``Why Hyperbolic Theories of Dissipation
Cannot be Ignored:  Comments on a paper by Kostadt and Liu", to appear
in Phys Rev D (2001)

\bibitem{WIJS79} W. Israel, J. Stewart, Ann. Phys. 118, 341 (1979)

\bibitem{LMR86} I.S. Liu, I. Muller, T Ruggeri, Ann Phys 169, 191 (1986)

\bibitem{MTW73} C.W. Misner, K.S. Thorne, J.A. Wheeler, "Gravitation"
(W.H. Freeman, New York, 1973), pp 471-483.

\bibitem{MR93} I. Muller, T. Ruggeri, Extended Thermodynamics (Springer-Verlag,
Berlin, 1993)

\bibitem{JS71} J. M. Stewart, ``Non-Equilibrium Relativistic Kinetic
Theory", (Springer-Verlag, NY, 1971)

\bibitem{CW79} C. M. Will, in "General Relativity:  An Einstein Centenary
Survey", S.W. Hawking, W. Israel, eds (Cambridge Univ Press, Cambridge), 
pp 24-89.

\end{thebibliography}
\end{document}